\documentclass{article}
\usepackage{spconf,amsmath,graphicx,hyperref, xcolor, amsfonts, amssymb, bm}
\usepackage[utf8]{inputenc}
\usepackage[T1]{fontenc}
\usepackage{cleveref}
\usepackage{orcidlink}

\newcommand{\orcid}[1]{\href{https://orcid.org/#1}{\textcolor[HTML]{A6CE39}{\aiOrcid}}}

\def\D\mathbf{D}
\title{Full band denoising of room impulse response in the wavelet domain with dictionary learning}
\name{Théophile Dupré \orcidlink{0009-0004-3662-1535}, Romain Couderc \orcidlink{0000-0002-8696-6361}, Miguel Moleron \orcidlink{0000-0003-1341-9997}, Axel Coulon, Rémy Bruno, Arnaud Laborie\thanks{}}
\address{Trinnov Audio, Neuilly-Plaisance, France}
\begin{document}
%
\maketitle
\begin{abstract}
Conventional wavelet-domain methods for room impulse response denoising rely on thresholding detail coefficients, which is unsuited for low frequencies. In this work, we introduce a wavelet-based post-processing algorithm that extends denoising to approximation coefficients by means of sparse dictionary learning with a time-varying error tolerance. The proposed method leverages an exponential decay envelope model to adapt reconstruction accuracy according to the local signal-to-noise ratio. This approach significantly improves low-frequency denoising of synthetic and measured room impulse responses compared to the baseline method, leading to more accurate estimation of acoustic parameters such as decay time.
\end{abstract}
\section{Introduction}
\label{sec:intro}


The room impulse response (RIR) characterizes the transfer function between a sound source and a receiver in a room. It inherently captures the acoustic properties of the source, the receiver, and the environment. RIRs are widely used in applications ranging from auralization and immersive technologies such as virtual and augmented reality \cite{vorlander2008auralization}, to the analysis of room acoustics \cite{weinzierl2018measuring}, and the calibration of reproduction systems \cite{cecchi2017room}.

The RIR can be obtained by exciting the system with a known signal, such as a maximum length sequence (MLS) \cite{schroeder1979integrated} or an exponential sine sweep \cite{farina2000simultaneous}, which covers the frequency range of interest, typically $20$Hz to $20$kHz in audio applications. The resulting signal is recorded at the receiver, and the RIR is then derived through a deconvolution process. As in any measurement procedure, the accuracy and reliability of the obtained RIR are subject to degradation in the presence of noise.

Low-frequency noise is especially critical in building environments, originating from sources such as ventilation systems or external vibrations transmitted through structural elements. Due to the reduced sensitivity of the human auditory system at low frequencies, these disturbances are often barely perceptible and thus difficult to detect during measurement. Moreover, they are challenging to suppress since they arise from factors external to both the room and the measurement procedure.

A variety of post-processing techniques have been proposed to mitigate noise in acoustic measurements. The most widely used approach consists in truncating the impulse response once noise prevails, followed by a compensation procedure \cite{ISO3382-1}, though this strategy cannot recover information beyond the truncated portion. In the context of multi-sensor measurements, correlation across sensors has been leveraged to reconstruct the missing signal segment \cite{masse2020denoising}. Methods inspired by speech enhancement, such as Fourier-domain spectral subtraction, have also been applied to RIRs \cite{chen2021noising}, but they require prior estimation of the noise spectrum. Other approaches inspired by image processing have explored wavelet-domain denoising of RIRs \cite{damnjanovivc2018noising}. Unlike fixed-resolution time–frequency representations, wavelet-based representations concentrate the signal energy into a small number of high-amplitude coefficients, while noise is distributed across many low-amplitude coefficients, which enables effective thresholding. However, these approaches are typically effective only above a selected cutoff frequency and therefore do not adequately process low-frequency components.

This paper introduces a post-processing denoising approach when noise-only samples are not available, specifically targeting low-frequency noise, while retaining the benefits of wavelet-based techniques to achieve effective noise reduction across the full audible spectrum. 




\section{Model}
\label{sec:model}

\begin{figure}
\centering
\includegraphics[width=0.95\columnwidth]{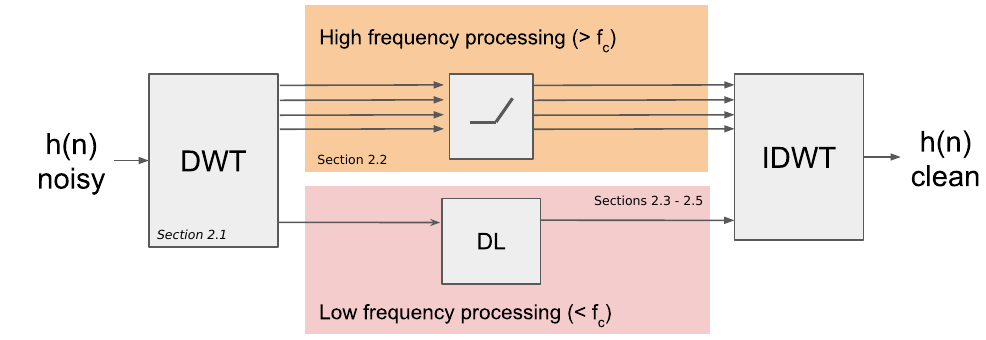}
\caption{Diagram of the proposed approach.}
\label{fig:schema}
\end{figure}

The proposed denoising algorithm builds on the wavelet-domain approach introduced by \cite{damnjanovivc2018noising}. The diagram in Figure \ref{fig:schema} outlines the workflow of the proposed denoising algorithm. It begins with the discrete wavelet transform (Section \ref{sec:dwt}), which decomposes the RIR into approximation and detail coefficients. The detail coefficients are denoised using thresholding (Section \ref{sec:threshold}), while the approximation coefficients are processed through dictionary learning (DL) with a time-varying error tolerance (Sections \ref{sec:dl}–\ref{sec:dl_env}). Finally, the inverse wavelet transform recombines the coefficients to reconstruct the denoised RIR.

\subsection{Discrete wavelet transform}
\label{sec:dwt}
Let $h[n]$, with $n \in  \{0, 1, \dots, N-1 \}$ sampled at frequency $f_s$, represent a measured RIR between a source and a receiver in a room. Using the discrete wavelet transform (DWT) \cite{mallat2002theory}, $h[n]$ can be decomposed into approximation coefficients $a_0[n]$ and detail coefficients $d_0[n]$:
\begin{equation*}
a_0[n] = \sum_{k=-\infty}^{+\infty} g[k]h[2n-k],
\
d_0[n] = \sum_{k=-\infty}^{+\infty} q[k]h[2n-k],
\end{equation*}
where $g[k]$ and $q[k]$ are quadrature mirror filters, acting respectively as low-pass and high-pass filters. These filters, derived from a mother wavelet (e.g., Haar, Daubechies, Meyer \cite{dautov2018wavelet}), define the type of decomposition, with the choice depending on the application.

The operations on $\mathbf{h}$ can be interpreted as convolution followed by downsampling by a factor of 2. As a result, $\mathbf{a_0}$ and $\mathbf{d_0}$ are sampled at $f_s/2$. This process constitutes the first-level analysis stage. Iteratively applying the same operation to the approximation coefficients yields multi-level decompositions, producing $\mathbf{a_1}, \mathbf{d_1}, \dots, \mathbf{a_{L-1}}, \mathbf{d_{L-1}}$ up to level $L$. The full set of coefficients $[\mathbf{d_0}, ..., \mathbf{d_{L-1}}, \mathbf{a_{L-1}}]$ provides a multi-resolution representation of the original signal $\mathbf{h}$, with frequency content localized at progressively lower effective sampling rates $f_s^l = f_s/(2^{l+1})$, for $l \in \{ 0, \dots, L-1 \}$. 

The following sections describe the denoising procedure applied to the wavelet coefficient set $[\mathbf{d_0}, \ldots, \mathbf{d_{L-1}}, \mathbf{a_{L-1}}]$, from which a cleaned version of $\mathbf{h}$ is obtained using the inverse wavelet  \cite{mallat2002theory}.

\subsection{Denoising the detail coefficients}
\label{sec:threshold}
A common denoising strategy for signals corrupted by full-band uncorrelated noise is to apply thresholding to the detail coefficients $[\mathbf{d_0}, ..., \mathbf{d_{L-1}}]$. In the wavelet domain, structured signal energy is concentrated in a small number of coefficients, while noise is spread more uniformly. A threshold is therefore applied to distinguish significant coefficients (assumed to contain signal) from low-energy coefficients (assumed to be noise), with the latter set to zero. Implementing thresholding requires selecting several parameters: how the threshold is estimated, how it is scaled, and which thresholding rule is applied. A detailed analysis of these factors can be found in \cite{damnjanovivc2018noising}.

However, a limitation of standard thresholding arises at low frequencies, since approximation coefficients $\mathbf{a_{L-1}}$ are not processed. To overcome this, we propose a DL approach designed to denoise the approximation coefficients, thereby extending denoising effectiveness into the low-frequency band. The cut-off frequency between the processing of high and low frequencies is then defined as the lowest effective sampling rate of the wavelet analysis : $f_c = f_s/2^{L}$.
\subsection{Dictionary learning}
\label{sec:dl}
For a downsampled signal of size $N_{L-1} = \frac{N}{2^L}$, the signal $\mathbf{a}_{L-1} \in \mathbb{R}^{N_{L-1}}$ may be modeled with the linear model~\cite{atomic_decomposition, conv_dict} as follows
\begin{equation}
    \mathbf{a_{L-1}} = \sum_{i=1}^{K}  z_i \boldsymbol{\delta_i}  + \boldsymbol{\xi},
\end{equation}
where $\boldsymbol{\delta_i} \in \mathbb{R}^{N_{L-1}}$ are $K$ unknown unit-norm atoms ($\|\boldsymbol{\delta_i}\| = 1$),  $z_i \in \mathbb{R}$ are $K$ unknown activation signals, and 
$\boldsymbol{\xi}$ is noise (not specifically white). In the literature, the set of atoms $\{\boldsymbol{\delta_i}\}_{i=1}^{K}$ is referred to as the \emph{dictionary}. The goal of DL is to jointly estimate the true activations $\bm{z}_i$ 
and atoms $\boldsymbol{\delta}_i$ from noisy observations $\bm{a}_{L-1}$. 
Various optimization formulations exist; in this work, we adopt the error-constrained sparse dictionary learning approach~\cite{Aharon2006_KSVD, Rubinstein2008_EfficientKSVDBatchOMP}. These methods construct a matrix $\mathbf{A}$ from the signal $\mathbf{a}_{L-1}$ using a sliding window of length $d$, where each column corresponds to a consecutive segment of $d$ samples shifted by one. The resulting Hankel-structured matrix $\mathbf{A} \in \mathbb{R}^{d \times (N_{L-1}-d)}$ is highly redundant. The objective is therefore to learn a dictionary $\mathbf{D}$ and a sparse activation matrix $\mathbf{Z}$ such that $\mathbf{A}$ can be accurately reconstructed by solving the following optimization problem:
\begin{equation}\label{prob_main}
\begin{aligned}
    \min_{\bm{Z} \in \mathbb{R}^{K \times (N_{L-1}-d) },\; \bm{D} \in \mathcal{C}} 
    & \quad \sum_{i = 1}^{N_{L-1} -d} \|\bm{Z}_i\|_0 \\
    \text{s.t.} \quad 
    & \|\bm{A}_i - \bm{D} \bm{Z}_i\|_2^2 \leq \varepsilon
\end{aligned}
\end{equation}
where $\mathcal{C} = \{\bm{D} \in \mathbb{R}^{d \times K} : ||\bm{D}_j||_2 \leq 1, \forall j \in \{1, \dots, K\} \}$, the operator $\| \cdot \|_0$ denotes the $\ell_0$ pseudo-norm (number of non-zero entries) and $\varepsilon > 0$ the error tolerance that controls the trade-off between error in the reconstruction and sparsity. This formulation is particularly relevant for RIR denoising, as it allows $\varepsilon$ to be adapted to the signal characteristics. In practice, we would like a sample-dependent tolerance: smaller errors at RIR samples with high SNR and larger errors where the SNR is low, ensuring tight reconstruction where the signal is reliable and more flexibility in noisier regions. For this purpose, it is necessary to have a fine estimation of the envelope of the RIR. 

\subsection{Envelope estimation}
\label{sec:env}
The decay envelope of a RIR in the presence of stationary noise 
can be modeled as an exponential decay plus a constant~\cite{Jankovic2016_RIR_Truncation_NonlinearDecay}:
\begin{equation}
    h[n] = x_1 e^{-x_2 n} + x_3, 
    \quad \forall n \in \{0, \dots, N-1\},
\end{equation}
where $\bm{x} = [x_1, x_2, x_3]^\top \in \mathbb{R}^3$ are the model parameters representing 
the initial level ($x_1$), the decay rate ($x_2$), and the stationary noise floor ($x_3$).  

The parameters $\bm{x}$ are estimated via nonlinear least squares by fitting the model 
to the measured envelope. The fitting problem is formulated as:
\begin{equation}\label{prob_envelop}
\small
    \min_{\bm{x} \in [\bm{b}_\ell, \bm{b}_u]}  
    \sum_{n = 0}^{N-1} \left[ 
        \log_{10}\!\left(h[n]^2\right) - 
        \log_{10}\!\left(x_1^2 e^{-2x_2 n} + x_3^2 \right) 
    \right]^2,
\end{equation}
where $\bm{b}_\ell, \bm{b}_u \in \mathbb{R}^3$ denote the lower and upper bounds 
on the parameters, respectively. These bounds prevent ill-posed solutions, 
for instance when the initialization is poorly chosen. 
Problem~\eqref{prob_envelop} is solved using the Levenberg–Marquardt algorithm~\cite{Marquardt1963}.

\subsection{Dictionary learning with a time-varying error}
\label{sec:dl_env}
Once Problem~\eqref{prob_envelop} is solved, the estimated envelope is used 
to compute a time-varying reconstruction error $\varepsilon[n] > 0$. 
This error is based on two quantities derived from the estimated envelope~\cite{Jankovic2016_RIR_Truncation_NonlinearDecay}:  
\begin{equation*}
\begin{array}{lll}
\text{Noise-to-signal ratio:}  & 
c_{\mathrm{nsr}} = \tfrac{x_3}{x_1} \\[6pt]
\text{Transition time:}  & 
T_t = \frac{-\log(x_3/x_1)}{x_2}.
\end{array}
\end{equation*}
The time-varying error is then defined as
\begin{equation*}
\varepsilon[n] = 
\begin{cases}
10^{-4}, & n \leq \lfloor T_t \rfloor, \\[4pt]
1 - \exp\!\left(-\tfrac{x_2}{f_s}(n - \lfloor T_t \rfloor)c_{\mathrm{nsr}}\right), & n > \lfloor T_t \rfloor .
\end{cases}
\end{equation*}
This error term reflects two complementary regimes.  Before the transition time $T_t$, the signal energy dominates the noise energy, and the error tolerance is kept very small to enforce precise reconstruction (the value $10^{-4}$ was chosen experimentally and found to work reliably). 
Beyond $T_t$, the signal energy decays exponentially while the stationary noise remains constant. In this region, the error tolerance increases progressively at a rate governed by the estimated decay $x_2$ 
and scaled by the NSR. A larger NSR leads to a steeper error growth because fewer reliable samples are available for reconstruction.

Finally, the optimization is solved iteratively by alternating between two steps: (i) \textbf{sparse coding}, where an Orthogonal Matching Pursuit (OMP) algorithm (Algorithm~1 in~\cite{Rubinstein2008_EfficientKSVDBatchOMP}) is applied to estimate $\bm{Z}$, with the stopping criterion determined by the time-varying error $\varepsilon[n]$; and (ii) \textbf{dictionary update}, where an estimated K-SVD algorithm (Algorithm~5 in~\cite{Rubinstein2008_EfficientKSVDBatchOMP}) is used to update $\bm{D}$. 

\section{Results}
\label{sec:results}
Numerical experiments were conducted on both simulated and experimental data, 
with the primary objective of enhancing the low-frequency denoising performance 
of the baseline method~\cite{damnjanovivc2018noising}. For this purpose, both methods are compared on a set of numerical experiments where low-frequency noise have been artificially added. For these experiments, the DWT was configured with the same parameters as in~\cite{damnjanovivc2018noising}, except that discrete Meyer wavelet
was used instead of Haar wavelet in order to achieve smoother decompositions, with a decomposition level set to $L = 8$. For the DL stage of the proposed method, the number of atoms was set to $K = 8$, and their length was chosen as $d = N_{L-1}/2$.

\subsection{Decay Time (DT) estimation}
The $\text{DT}_{60}$ is defined as the time required for the sound pressure level in a room (or in an impulse response) to decay by $60$ dB after the source has stopped. At low frequencies, it is more appropriate than reverberation time because the acoustic field is not diffuse.
However, estimated $\text{DT}_{60}$ in the presence of noise, especially in low frequencies, is challenging \cite{MEISSNER_2013} because the noise floor masks the latter part of the decay. This generally leads to an overestimation of the value of $\text{DT}_{60}$ which can be detrimental in practice. 

\begin{figure}[h!]
    \centering
    \includegraphics[width=1\linewidth]{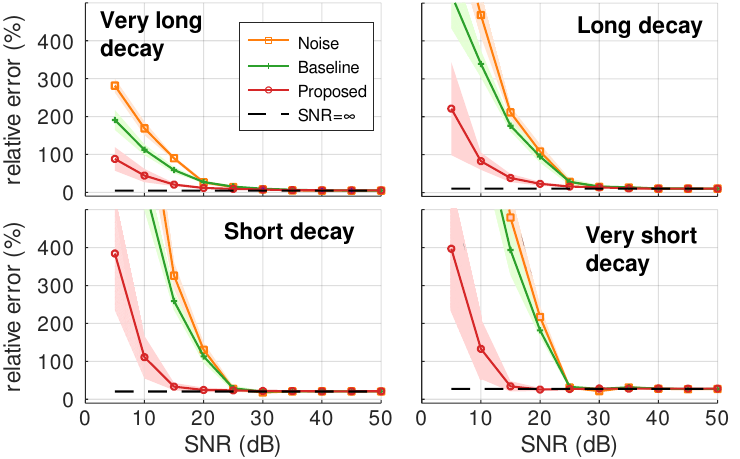}
    \caption{Estimated $\text{DT}_{60}$ of both methods for different levels of noise and attenuation.}
    \label{fig:simu_decay}
\end{figure}

To evaluate the effectiveness of the proposed method, a set of test signals was generated as
\begin{equation*}
    h[n] = \sum_{k = 1}^K s_k \, e^{-\alpha_k n} \, \sin(\omega_k n), 
    \quad n \in \{0, \dots, N-1\},
\end{equation*}
where $s_k$ and $\alpha_k$ denote the amplitude and decay factor of the $k$-th sinusoid, respectively, and $\omega_k$ is its angular frequency, corresponding to the center frequency of the $k$-th third-octave band. The frequencies span the range from $25$~Hz to $100$~Hz. This modal representation is particularly well suited for modeling low-frequency signals and allows exact calculation of the decay time $\text{DT}_{60}^k = \frac{3}{\alpha_k\log_{10}(e)}$ for each signal. 

Ten different filtered white noises, shaped to match the noise spectrum observed in experimental measurements, were then added to the signals at ten equally spaced SNR levels from 5 to 50~dB. Finally, the experiment was repeated four times, with the initial decay multiplied by a factor $f \in \{0.5, 1, 1.5, 2\}$.

The results are presented in \Cref{fig:simu_decay}. As observed in~\cite{MEISSNER_2013}, the $\mathrm{DT}_{60}$ can be reliably estimated 
for SNRs between 25 and 35~dB. Below this range, the estimation error of both the noisy signal and the baseline method increases rapidly. In contrast, the proposed method maintains relatively low estimation error down to an SNR of 15~dB, particularly for smaller decay rates, demonstrating its effectiveness in denoising the low-frequency band.
\subsection{Experimental results}
\begin{figure}[h!]
    \centering
    \includegraphics[width=\columnwidth]{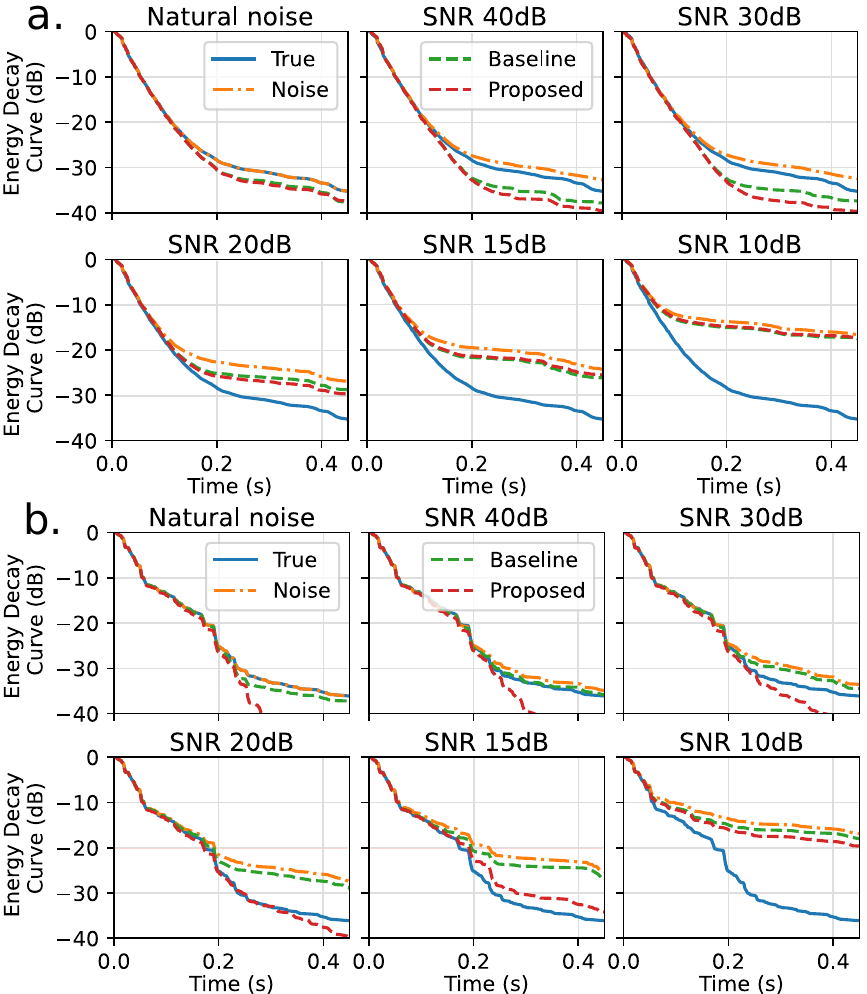}
    \caption{Schroeder's integral for different levels of noise on signal from \textbf{a.} large band speakers and \textbf{b.} subwoofers.}
    \label{fig:schroder_int}
\end{figure}
In practice, the exact $\mathrm{DT}_{60}$ of a room cannot be known. Therefore, to evaluate the effectiveness of the proposed method on real measurements, we rely on \emph{Schroeder's integration} \cite{Schroeder1965}, a standard technique in room acoustics to estimate the energy decay curve from a measured RIR. As explained in~\cite{damnjanovivc2018noising}, denoising is considered effective if the energy decay curve of the noisy measurement closely follows that of the clean signal without falling below it, since a dip below the clean curve would indicate loss of signal. In our experiments, two large-band loudspeakers and four subwoofers were measured at ten different positions, with noise levels generated in the same way as in the simulated experiments and artificially added via an additional loudspeaker.
\begin{figure}[h!]
    \centering
    \includegraphics[width=1\linewidth]{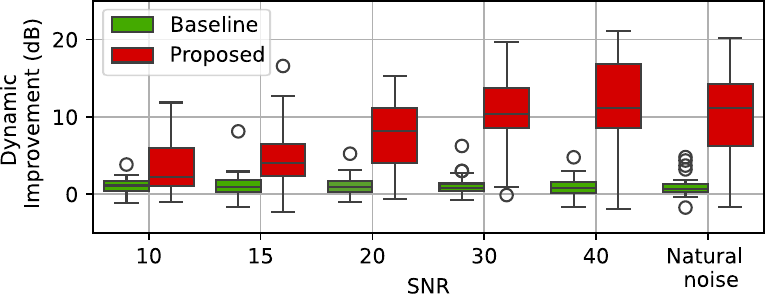}
    \caption{Comparison of dynamic range improvement between the baseline and the proposed method across different SNR levels.}
    \label{fig:dyn_imp}
\end{figure}

 An instance of the results for a large band speaker and a subwoofer is presented in \Cref{fig:schroder_int} a and b, respectively. 
 In these plots, the true curve corresponds to the measurement without artificially added noise, while still including the natural background noise. The noisy curve includes both natural and artificially added noise. In low-SNR cases, the proposed method yields estimates that are closer to the true curve than the baseline, demonstrating superior performance in removing the artificially added noise. In high-SNR cases, the proposed method is able to extrapolate the decay further, indicating that it also suppresses the natural noise present in the room during the experiment.
 Nevertheless, the proposed method performs less effectively for the large-band speaker (\Cref{fig:schroder_int}a). This behavior can be attributed to the weaker low-frequency content emitted by this speaker, which makes denoising more challenging.

Another metric, also used in \cite{damnjanovivc2018noising}, is the \emph{dynamic improvement}, which is defined as the noise floor level difference before and after processing. Figure \ref{fig:dyn_imp} presents the dynamic improvement for the set of experimental measures and it consistently outperforms the baseline that struggles to improve the dynamic in the presence of low frequency noise.

\section{Conclusion}
\label{sec:conclusion}

This work introduced a wavelet-based post-processing method that extends denoising to low frequencies through sparse DL with adaptive error control. The proposed approach outperforms the baseline, yielding more accurate RIR measurements and improved estimation of acoustic parameters especially in low frequencies. Further works would be to investigate the learned sparse representation which may contain information about the room, the source or the receiver.

\vfill\pagebreak

\bibliographystyle{IEEEbib}
\bibliography{strings,refs}

\end{document}